# A multi-institutional pediatric dataset of clinical radiology MRIs by the Children's Brain Tumor Network

Ariana M. Familiar[1,2], Anahita Fathi Kazerooni[1,2,3], Hannah Anderson[1,4], Aliaksandr Lubneuski[1,2], Karthik Viswanathan[1,2], Rocky Breslow[1,2], Nastaran Khalili[1,2], Sina Bagheri[1,4], Debanjan Haldar[1,2], Meen Chul Kim[1,2], Sherjeel Arif[1,2], Rachel Madhogarhia[1,2], Thinh Q. Nguyen[1,2], Elizabeth A. Frenkel[1,2], Zeinab Helili[1,2], Jessica Harrison[1,2], Keyvan Farahani[5], Marius George Linguraru[6,7], Ulas Bagci[8], Yury Velichko[8], Jeffrey Stevens[9], Sarah Leary[9], Robert M. Lober[10], Stephani Campion[11], Amy A. Smith[11], Denise Morinigo[12], Brian Rood[12], Kimberly Diamond[13], Ian F. Pollack[13], Melissa Williams[14], Arastoo Vossough[1,4,15], Jeffrey B. Ware[1,4], Sabine Mueller[16], Phillip B. Storm[1,2], Allison P. Heath[1,2], Angela J. Waanders[14,17] , Jena V. Lilly[1,2], Jennifer L. Mason[1,2], Adam C. Resnick[1,2], Ali Nabavizadeh[1,4]*

[1] Center for Data-Driven Discovery in Biomedicine, Children's Hospital of Philadelphia, Philadelphia, PA, USA

[2] Department of Neurosurgery, Children's Hospital of Philadelphia, Philadelphia, PA, USA

[3] Department of Neurosurgery, Perelman School of Medicine, University of Pennsylvania, Philadelphia, PA, USA

[4] Department of Radiology, Perelman School of Medicine, University of Pennsylvania, Philadelphia, PA, USA

[5] National Cancer Institute, Bethesda, MD, USA

[6] Sheikh Zayed Institute for Pediatric Surgical Innovation, Children's National Hospital, Washington, DC, USA

[7] Departments of Radiology and Pediatrics, George Washington University School of Medicine and Health Sciences, Washington, DC, USA

[8] Department of Radiology, Feinberg School of Medicine, Northwestern University, Chicago, IL, USA

[9] Department of Hematology and Oncology, Seattle Children's, Seattle, WA, USA

[10] Division of Neurosurgery, Dayton Children's Hospital, Dayton, OH, USA

[11] Department of Pediatric Hematology & Oncology, Orlando Health Arnold Palmer Hospital for Children, Orlando, FL, USA

[12] Department of Hematology-Oncology, Children's National Hospital, Washington, DC, USA

[13] Department of Pediatric Neurosurgery, UPMC Children's Hospital of Pittsburgh, Pittsburgh, PA, USA

[14] Division of Hematology, Oncology, NeuroOncology, and Transplant, Ann & Robert H Lurie Children's Hospital of Chicago, Chicago, IL, USA

[15] Department of Radiology, Children's Hospital of Philadelphia, Philadelphia, PA, USA

[16] Department of Neurology, Division of Child Neurology, University of San Francisco, San Francisco, CA, USA

[17] Department of Pediatrics, Northwestern University Feinberg School of Medicine, Chicago, IL, USA

* corresponding author: Ali Nabavizadeh ali.nabavizadeh@pennmedicine.upenn.edu

**Abstract**

Pediatric brain and spinal cancers remain the leading cause of cancer-related death in children. Advancements in clinical decision-support in pediatric neuro-oncology utilizing the wealth of radiology imaging data collected through standard care, however, has significantly lagged other domains. Such data is ripe for use with predictive analytics such as artificial intelligence (AI) methods, which require large datasets. To address this unmet need, we provide a multi-institutional, large-scale pediatric dataset of 23,101 multi-parametric MRI exams acquired through routine care for 1,526 brain tumor patients, as part of the Children's Brain Tumor Network. This includes longitudinal MRIs across various cancer diagnoses, with associated patient-level clinical information, digital pathology slides, as well as tissue genotype and omics data. To facilitate downstream analysis, treatment-naïve images for 370 subjects were processed and released through the NCI Childhood Cancer Data Initiative via the Cancer Data Service. Through ongoing efforts to continuously build these imaging repositories, our aim is to accelerate discovery and translational AI models with real-world data, to ultimately empower precision medicine for children.

## Background & Summary

For brain cancer patients, radiology images are routinely collected and used to inform decision making across a variety of applications including surgical planning, longitudinal tumor monitoring, and assessment of treatment response. Consequently, a significant amount of imaging data (primarily MRIs) is generated through clinical standard of care. Over the past decade there have been several large-scale, multi-institutional initiatives to collect and curate such data for adult populations and make them publicly available for research purposes. For example, The Cancer Imaging Archive (TCIA) provides several collections focused on adult glioblastoma multiform (GBM)[1,2,3,4]. These efforts have contributed to the empowerment of predictive artificial intelligence (AI) and machine learning (ML) methods in cancer research, which require ample amounts of representative data that capture variance in image acquisition protocols and hardware at different institutions. In the context of adult brain cancer research, there have been a significant number of studies that examine "radiomic" features -- the high-throughput image-derived properties of a tumor-affected region -- and have used ML and deep learning based approaches to use radiomic features to predict other patient-level factors such as molecular genotype (radiogenomics) or clinical outcome measures such as survival prognosis or response to therapy[5,6]. The translation of trained AI models into clinical practice workflows has also started to become a reality[7,8], and although continued work is required to validate and generalize existing models, this indicates a crucial step towards harnessing the predictive power of AI for precision medicine in neuro-oncology[9,10].

On the other hand, the availability of clinical MRIs that longitudinally and comprehensively capture a patient's treatment journey, beyond just treatment-naïve timepoints, remains elusive. The lack of publicly accessible longitudinal scans has limited the development of radiomic models that can account for post-treatment imaging changes (such as areas of necrosis, post-resection gliosis, and edema), which are crucial for assessment of treatment response. Additionally, many public datasets focus on a single cancer type (such as GBM), and do not include other less frequent types of brain tumors that would be essential for building diagnostic tools that can perform optimally for different tumor histologies. Recently, in 2023, the Brain Tumor Segmentation challenge (BraTS) has expanded its scope to include not only primary brain tumors but also brain metastases, meningiomas, and other related conditions. This expansion reflects the growing recognition of the importance of comprehensive analysis and segmentation in various types of brain tumors. In the context of pediatric brain tumors (PBTs), there is a dearth of AI models that are able to generalize beyond the context of a single study, largely due to a lack of data availability[11]. PBTs are relatively rare compared to adult populations but remain one of the leading causes of cancer-related mortality in children. Moreover, brain cancers in adults often have differing radiological, molecular, and clinical characteristics compared to pediatric cases; as a result, existing models that have been trained on adult data do not typically perform well on pediatric data. Aggregating MRIs across institutions can accelerate translational research on PBTs, particularly for less common cancer histologies, although cross-site collection is made difficult by data privacy regulations and technical infrastructure requirements to ensure secure data transfer. To date, only one pediatric radiology dataset has been made available via the TCIA[12], which was from a Children's Oncology Group clinical trial (NCT00392327) that focused on high-risk medulloblastoma, supratentorial primitive neuro-ectodermal tumor of the CNS (CNS-PNET), and pineoblastoma (PBL)[13]. This includes pre- and post-operative MRIs of the brain and spine, with and without contrast, for 85 subjects.

To address the aforementioned challenges, we made available a centralized large dataset of clinical MRIs as part of the Children's Brain Tumor Network (CBTN)[14]. The CBTN is an international consortium of 34 healthcare institutions that contribute to a multi-modal data repository as well as collaborative research efforts focused on pediatric neuro-oncology funded in large part by philanthropic organizations. As of March 2023, CBTN has enrolled over 4,900 subjects under a shared regulatory protocol that allows centralization of patient-level data and release to the wider community through NIH-supported software platforms such as the Gabriella Miller Kids First Data Resource and the National Cancer Institute (NCI) Childhood Cancer Data Initiative (CCDI). This has allowed the creation of CBTN's Pediatric Brain Tumor Atlas (PBTA)[15] for which over 3,000 surgically collected tumor specimens have been released with paired clinical data (demographic, treatment, outcome). Many PBTA samples have associated whole genome sequencing (WGS), tumor RNA-seq, and digital histopathology images, and some have methylation and/or proteomic data. Recent grant funding from the Kids First Pediatric Research Program (X01) has initiated the molecular sequencing of an additional 4,594 tissue samples, including germline and somatic variants. This initiative has established an unprecedented multiomic dataset to support advancements in the study and treatment of PBTs. Through joint CBTN efforts, we have collected 23,101 imaging exams for over 1,500 subjects, which is freely available via secure access mechanisms in alignment with FAIR principles (Findable, Accessible, Interoperable, and Reusable) for data stewardship[16]. Images were acquired through institution-specific clinical imaging protocols as part of routine standard of care, and acquisition protocols across institutions were not uniform. Crucially, as opposed to images collected as part of research protocols, this radiological dataset captures real-world heterogeneity in image types and acquisition parameters that vary across scanners and sites and as such, it can facilitate the generalizability of downstream radiomic and integrated multiomic AI models, which has been a challenge across both adult and pediatric contexts to-date. The CBTN dataset is comprised of longitudinal MRIs across treatment-naïve, post-surgical, systemic treatment, and follow-up timepoints, as well as at events of progression or relapse, and includes subjects across a variety of cancer histologies. All subjects with MRIs have associated clinical data, including demographics, treatment information, and survival outcomes, as well as digital pathology slides (whole slide images, WSIs). Many subjects also have genomic and other -omic data, making the dataset is suitable for use in radiomic, radiogenomic, radiopathomic, and other integrative multi-omic analyses. Notably, we will continue to collect MRIs for CBTN subjects to iteratively grow the dataset and will freely provide all centralized data to the broader research community.

To further prepare the imaging data for research contexts and enable its rapid use, we selected a subset of subjects with available treatment-naïve timepoints that had four main multi-parametric scans (T1-weighted, T1-weighted contrast-enhanced, T2-weighted, and T2-FLAIR) and processed them using standard methods. We provide the pre-processed images and brain masks (with removed face features, "defaced") as a secure access CCDI collection through the NCI Cancer Research Data Commons (CRDC) via the Cancer Data Service (CDS; https://datacommons.cancer.gov/repository/cancer-data-service), with direct association to clinical and genomic CBTN data stored in the CDS. These four particular image sequences are relevant across many applications, including: (1) tumor region delineation, as in existing BraTS datasets[17,18]; (2) response assessment, such as defined by the Response Assessment in Pediatric Neuro-Oncology working group[19–22]; and (3) are commonly used in radiomic studies[23]. The motivation for this effort is to reduce the time-consuming, manual burden and the technical and domain-expertise required to prepare images for downstream analysis, and thus allow maximum utility of the data in research settings. By providing these resources, we aim to accelerate the development of predictive analytics with radiology imaging in pediatric neuro-oncology and advance the application of resultant models into clinical decision-support workflows.

The CBTN imaging data described in this paper is released in two separate locations, each with their own set of access procedures and data organization standards (Table 1; both are provided in NIfTI file format and are de-identified of protected health information). Where appropriate, we outline separately the "CBTN-Flywheel" dataset, which contains the entirety of the collected imaging exams that are provided in their "raw" or unprocessed format (i.e., without any image-based processing following retrieval from clinical radiology systems), and the "NCI-CDS" dataset, which contains the selected set of pre-processed, treatment-naïve images that are additionally "de-faced" (facial regions removed).

## Methods

### *Data collection and de-identification*

A CBTN master agreement allows the Center for Data-Driven Discovery in Biomedicine (D3b) at the Children's Hospital of Philadelphia (CHOP, Philadelphia, PA; Figure 1) to act as the Operations and Data Coordinating Center for the CBTN. This includes oversight of the sample and data management (including files containing protected health information; PHI) and sharing of de-identified samples/data by CHOP with CBTN members and CBTN-approved researchers following approval by the CBTN Scientific Committee (see *Usage Notes* for more details on the request process). For controlled access data, such as imaging data, the master agreement states that the release of data to researchers is allowed under a signed Data Use Agreement (DUA). The CBTN DUA terms were drafted in compliance with NIH guidelines and applicable laws, and its purpose is to facilitate the release of data collected across healthcare sites while also protecting all involved parties, including patients, patient families, and member institutions. During onboarding of a site to the CBTN consortium, the master agreement is reviewed, approved, and signed by each site, including review by their site-specific legal team. The master agreement states that individual sites are responsible for properly consenting each subject prior to sharing their data with CHOP and the CBTN. In addition, each of the 34 CBTN sites has their own site-specific regulatory documents, which are approved by their Institutional Review Board (IRB) and are reviewed by the CHOP Operations team to ensure key language is included that agrees with the terms in the master agreement. Sites are also able to utilize a CBTN template IRB protocol and consent documentation for such purposes. CHOP maintains records of all sites' regulatory documents and updates them as they expire each year.

MRIs were collected across 7 contributing sites: CHOP, Seattle Children's Hospital (Seattle, WA), Lurie Children's Hospital of Chicago (Chicago, IL), Dayton Children's Hospital (Dayton, OH), Orlando Health Arnold Palmer Hospital (Orlando, FL), Children's National Hospital (Washington, DC), and UPMC Children's Hospital of Pittsburgh (Pittsburgh, PA). The selection criteria included MRI exams of the brain and/or spine but could include additional examined body parts if feasible for collection. Corresponding DICOM files were retrieved from each site's Radiology Department PACS (Picture Archiving and Communication System) system using site-specific request processes and sent to CHOP via available imaging transfer platforms (Ambra, Powershare). The majority of transferred MRIs were identified (containing PHI) as permitted by the CBTN master agreement.

There were no selection criteria based on clinical factors. MRIs were collected across a variety of PBT histological diagnoses (confirmed via pathology free text reports and electronic health record (EHR) review by trained clinical coordinators and research assistants). These included low-grade glioma/astrocytoma (LGG), high-grade glioma/astrocytoma (HGG), medulloblastoma, ependymoma, atypical teratoid/rhabdoid tumors (ATRT), craniopharyngioma, diffuse intrinsic pontine glioma (DIPG), dysembryoplastic neuroepithelial tumors (DNET), neurofibroma, schwannoma, and other histologies (Fig. 2; see full list in Table 1).

DICOM files were obtained and de-identified (stripped of PHI) using processes developed in-house (described below) that have been made publicly open source (see *Code Availability* for information on shared resources). First, "Structured Report document" (SR) and "Other" (OT) modality acquisitions were removed (based on the DICOM Modality tag (0008,0060)). Acquisitions with a Series Description tag (0008,103E) that included any of the following text (case-insensitive) were removed: screensave/screen save/screen_save, cover image/cover_image, dose_report/dose report/dosereport, documents, protocol, capture. Studies with a Study Description tag (0008,1030) including the following text were removed: script, bone scan. DICOMs were then converted into NIfTI format (dcm2niix[24]). A BIDS[25] option (flag "-b") with anonymization was used to output a paired JSON file to be released with each NIfTI file that included a set of DICOM metadata fields not retained in NIfTI file headers. Additional fields were removed from the JSON to comply with DICOM-NEMA standards for PHI anonymization (Full list in NEMA guidelines Table E.1-1: https://dicom.nema.org/medical/dicom/current/output/html/part15.html#table_E.1-1), which included: Device Serial Number, Image Comments, Institution Address, Institutional Department Name, Institution Name, Procedure Step Description, Protocol Name, Station Name. Diffusion derivative files (gradient b-values/BVAL and directional b-vectors/BVEC) were retained. Lastly, acquisitions were removed if they did not have sufficient image acquisition DICOM tags in their JSON files, which removed any remaining files with "burned-in" PHI such as screenshots and scanned paperwork forms, as well as derivative files generated at the scanner such as region-of-interest images and images of physiological measurements, which were considered as non-interest. The resulting metadata and images of a subset of the exams were manually inspected to ensure comprehensive PHI removal. The temporal relationships between longitudinal acquisitions were retained by labelling each timepoint by a subject's age at the time of imaging (in days; see *Data Records*).

*Cohort selection and image pre-processing for the NCI-CDS dataset*

A subset of 370 subjects/sessions were selected for further processing based on availability of treatment-naïve imaging and four main sequences: T1-weighted (T1w), T1w with contrast enhancement (T1w-CE), T2-weighted (T2w), and T2-Fluid attenuated inversion recovery (FLAIR) scans. This included: 208 LGG, 99 medulloblastoma, 29 HGG, 18 DIPG, 8 ependymoma, 2 ETMR, 2 germinoma, 1 craniopharyngioma, 1 neurocytoma, 1 supratentorial PNET, and 1 teratoma (Table 1).

The four selected scans (T1w/T1w-CE/T2w/FLAIR) from a given treatment-naïve exam were pre-processed using the BraTS pre-processing pipeline toolkit released through the CaPTk software[28,29]. In brief, images were oriented to LPS/RAI coordinates (Left, Posterior, Superior / Right, Anterior, Inferior), and rigid registration was used to spatially register T1w/T2w/FLAIR to T1w-CE and all 4 scans to the SRI-24 atlas[30]. Images were resampled to a common resolution of 1 x 1 x 1 $mm^3$. Skull-stripped brain masks were generated using a pre-trained, pediatric-specific brain extraction deep learning model[31].

The four pre-processed scans were defaced using the FreeSurfer MiDeFace software package ("Minimally Invasive DeFacing")[26], which scrubs only surface-level voxels of facial regions in an image, in order to prevent reconstruction of a subject's face which can be identifiable. Each defaced image was visually inspected to ensure accurate and comprehensive defacing performance. If an image was not successfully defaced (typically in young children or due to low image resolution, and more commonly for T2w and FLAIR sequences as MiDeFace was developed based on T1w images), the image was manually defaced using the ITK-Snap software platform[27] (5% of images).

## Data Records

*CBTN-Flywheel (unprocessed data)*

The full, de-identified CBTN dataset (1,526 subjects, 23,101 exams) is made freely available to the community with secure access by the CBTN consortium and is managed by the CHOP Data Coordinating Center under the CBTN's master agreement terms. The dataset is provided as NIfTI files (with paired JSON DICOM metadata files) on CBTN's Flywheel platform. These data files are unprocessed, or "raw", in that there is no image-based processing of the data after collection from clinical radiology systems. Because image acquisition protocols were not uniform across sites, there is heterogeneity in scanner and image acquisition properties across exams. To receive access to the full dataset, researchers must complete a CBTN data request process (see instructions and details in *Usage Notes*).

Data is hosted on the CBTN's Flywheel website (www.chop.flywheel.io) and is organized in a hierarchical manner. CBTN subjects are grouped into unique "project" directories according to their diagnosis at the time of imaging (project are labelled by diagnosis, such as "Ependymoma"). Within a project, data is organized by subject and session (imaging exam/study). Users can download entire projects, or a selected subset of subjects or sessions.

Subject labels are unique, anonymized identifiers (CBTN Subject IDs) that are identical across data types to allow for multi-modal comparisons. Imaging session labels are generated based on DICOM tags and are formatted to include: age in days at imaging, primary body site examined, and acquisition time (hhmm); e.g., "100d_B_brain_12h30m". This enables unique session labels at a subject-level, while retaining longitudinal acquisition information without the use of any PHI-containing fields. Source images (acquisitions) are provided in compressed NIfTI file format. File names are derived from the Series Description (0008,103E) DICOM tag of the given acquisition (e.g., "t1_mprage_sag_p2_iso_0.9.nii.gz"). Each NIfTI file has a paired JSON file to provide additional DICOM metadata that is not retained in the NIfTI header.

Additional classification tags describing the image types were associated with each NIfTI file using a heuristic-based method. In this step, mappings between DICOM SeriesDescription substrings and uniform labels were established based on the entire dataset, and a corresponding dictionary was created. The final dictionary was used to classify all images using text-based pattern search against their SeriesDescription. This resulted in image-level tags describing the type of sequence, which can be used to query and filter the dataset according to specific parameters of-interest. This step was critical to ensure useability of the dataset because scans of the same type can have various SeriesDescriptions within and between scanners and institutions. To operate at-scale, this non-standard variance in clinically acquired images must be captured to enable flexible and comprehensive data curation.

*NCI-CDS (processed data)*

The subset of de-identified, processed treatment-naïve MRIs and brain masks (N=370; defaced, pre-processed images) are freely provided as a secure access dataset as part of the controlled access CBTN study released via NCI's data and platform ecosystem of the Cancer

Research Data Commons (CRDC) and Childhood Cancer Data Initiative (CCDI). The CBTN study is accessible through the NCI's Cancer Data Service (CDS) via the database of Genotypes and Phenotypes (dbGaP[32]; dbGaP study accession number for CBTN: phs002517; http://www.ncbi.nlm.nih.gov/projects/gap/cgi-bin/study.cgi?study_id=phs002517.v1.p1 ). To obtain access to this dataset, investigators must complete the dbGaP authorized access request process (see instructions and details in *Usage Notes*).

The NCI-CDS dataset is comprised of processed T1w/T1w-CE/T2w/T2w-FLAIR images that have been spatially co-registered and re-sampled, as well as corresponding deep learning-generated brain mask segmentations (see details in *Methods*). All files are provided as compressed NIfTI format and are organized by subject label (anonymized CBTN Subject IDs) and session label (age in days at imaging). Files are named based on the following naming convention: [Subject-ID]_[age]_[image type]_to_SRI_defaced.nii.gz ; where image type field includes one of "T1", "T1CE", "T2", "FLAIR", and "pred_brainMask", according to the image. Imaging data can be associated with subject-level genomic and molecular data using a mapping manifest of subject IDs and sample IDs (maintained and provided by dbGaP).

**Technical Validation**

Given that the dataset consists of MRIs acquired across scanners and sites, we aimed to further assess the quality of the constituent images to verify their use in research settings. For assessing general patterns of image acquisition properties in the full dataset, we extracted image resolution (voxel size) separately for all T1w (including T1w-CE), T2w, and FLAIR images as well as repetition (TR) and echo (TE) times for T1w and T2w images (Fig. 2E). With the exception of a few outlier images, the distribution of voxel sizes (T1w: x/y 0.14 – 2.0 mm,

median=0.75; z 0.4 – 34.8, median=3.3; T2w: x/y 0.27 – 4.1, median=3.13; z 0.4 – 30.0, median=10; FLAIR: x/y 0.2 – 1.72, median=0.69; z 0.5 – 24.5, median=4.0), TRs (T1w: 0.003 – 9.39, median=1.9; T2w: 0.004 – 15.5, median=2.5), and TEs (T1w: $7e^{-05}$ – 0.55, median=0.003; T2w: 0.002 – 0.53, median=0.014) fall within an acceptable range.

The unprocessed images of a subset of the selected pre-processed cohort (N = 150) were analyzed with the MRIQC package[33], which provides spatial image quality metrics (IQMs) for quality control based on the Quality Assessment Protocol (QAM). The tool was developed based on normal brain anatomy and has been tested on the large multi-site ABIDE (Autism Brain Imaging Data Exchange) dataset[34,35]; however, the influence of solid tumor presence on MRIQC measures has not been empirically determined. In particular, metrics that are calculated separately for white matter (WM), gray matter (GM), and CSF depend on accurate segmentation of these regions, which within the tool is performed with the FSL[36] (FMRIB Software Library) FAST[37] (FMRIB's Automated Segmentation Tool) software package after skull-stripping with AFNI[38] (Analysis of Functional NeuroImages) software packages. These tools are not likely to perform well when the brain anatomy has been impacted by tumor growth and mass effect on surrounding neural structures, as they were developed primarily for use in healthy individuals. Nonetheless, investigating the IQMs for global properties can provide a general characterization of image quality and can be used to examine the distribution of measurements across subjects within the multi-site dataset. This can be particularly useful for research studies on the harmonization of images acquired across different sites and scanners, as it captures real-world variance in relevant properties.

Thirteen standard (defined in Table 2) and four summary IQMs (mean, standard deviation, 5$^{th}$ and 95$^{th}$ percentiles and kurtosis) were extracted for T1w, T1w-CE, T2w, and FLAIR sequences separately (Fig. 3). T1w-CE and FLAIR results are included in the visualized plots for comparison, however MRIQC was designed for use on structural T1w and T2w images, so we only focus on the results for these image types. For T1w and T2w images, the mean of intensity values within white matter across subjects is centered around 1000 and within background is near 0 with little spread. The mean within gray matter is just under 1000 on T1 and 2000 on T2. Overall, the T1w results are consistent with prior literature[39], but the T2w values are higher than would be expected. This could be due to the bright appearance of tumor lesions on T2w (hypointense signal value) but the more moderate intensity values of the same lesion on T1w compared to normal white or gray matter, which would influence the summary statistics if tumor regions were included in the gray and/or white matter segmentations. The mean within CSF on T2w shows a wide range across subjects centered around 3500, and a smaller range on T1w centered around 500. Kurtosis and standard deviation are low on both T1w and T2w within white matter, gray matter, and CSF. In sum, these results show relatively small differences across subjects indicating comparable image intensity values within the dataset.

Several IQMs relate to noise and artifact measurements and can be used to identify image distortions caused by factors such as head motion or hardware-related magnetic field inhomogeneities. Across these measures, values across subjects indicate low presence of such artifacts (Table 2; Fig. 3; low average CJV, EFC, INU, QI1, QI2). Estimated blur across the image (FHWM) was also low (T1w in mm: $M$= 3.77, $SEM$=0.06; T2w: $M$= 4.06, $SEM$=0.1). It is important to note that the pre-processed cohort consists of subjects whose images were selected because they were free of obvious artifact, so the MRIQC metrics for this subset will not be representative of the presence of such image degradation in the full dataset. On the other hand, the metrics can be useful for assessment of variance based on images that were determined to be suitable for research-use, which can nonetheless capture batch effects caused by inter-site variability.

General IQMs reflecting signal intensities within the head compared to background (air surrounding the head; BG) showed acceptable results. Signal-to-noise (SNR) and contrast-to-noise (CNR) ratios were within the range of values of the $5^{th}$ - $95^{th}$ percentiles derived from 51,113 T1w samples and 767 T2w samples in an MRIQC crowdsourced database (Table 2; calculated based on data from "T1w_curated.csv" and "T2w_curated.csv" provided in prior study).[40]

IQMs comparing the volume of predicted WM, GM, and CSF indicate low agreement with standard measurements based on normal brain anatomy. For both T1w and T2w, estimated intracranial volume (ICV) of CSF (18% based on T1w) was comparable to the MRIQC benchmark (20%), but WM ICV was lower (36% vs. 45%) and GM ICV was higher (47% vs. 35%). This could either be due to incorrect segmentation maps in the brain tumor subjects (i.e., inaccurate tissue segmentation by the tool), or the comparison of pediatric and adult GM and WM volumes which are known to differ (GM decrease and WM increase in late childhood and adolescence[9]/29/23 12:54:00 PM). The tissue probability maps (TPMs) showed low overlap between WM, GM, and CSF maps of the images in the dataset and those derived from a common template based on a population atlas which could also be due to developmental age differences (the template is based on adult anatomy) and/or structural effects caused by the solid tumor on surrounding structures.

## Usage Notes

### *Public exploration portals and tools*

Users are able to explore the CBTN-Flywheel dataset properties, including imaging properties, clinical factors, and molecular data availability, without any approval necessary at the public website: https://d3b-rstudio-connect-public-prd.d3b.io/d3b-imaging-data-metrics/ . Detailed clinical and molecular data generated by CBTN can be explored on the public data portal website pedCBioPortal (https://pedcbioportal.kidsfirstdrc.org/ ; "PBTA" and "OpenPBTA" projects) and can be accessed with the same the CBTN research project request in the below-described request process. Patient IDs associated with each exam in the radiology dataset are identical to the Patient ID of the same subject with demographic, clinical, histopathological, and/or genomic data.

*Accessing the CBTN-Flywheel dataset*

A research project request form must be submitted (https://cbtn.org/research-resources; https://redcap.chop.edu/surveys/?s=A7M873HMN8) and a CBTN Data Use Agreement (DUA) must be signed. The request form includes details on who is requesting access to the data including project collaborators, what data types are being requested, and a Research Use Statement with a non-technical summary of the objective of the research project and how the data will be used and analyzed. Clinical report documents (including redacted pathologist, surgical, and radiologist notes), raw or processed genomic data, and/or access to digital pathology slides can also be requested through the same process. Requests will be reviewed to ensure consistency with scientific data use limitations, failure to include sufficient detail to determine this will result in rejection of a data request.

After approval, users are granted accessed to the dataset on the Flywheel platform (www.chop.flywheel.io; accessible in any standard web browser). Additionally, after project approval the CBTN Operations Team will provide requesters with all corresponding clinical data (e.g., demographics, medical history, diagnosis, treatment, survival outcomes) for the associated data. Requesters are also asked to provide project updates, and are suggested to return any derived results for further integration with the source dataset so as to reduce siloed research efforts and promote FAIR[16] practices. Moreover, researchers are requested to include the following acknowledgement statement to reference the data: "This research was conducted using data and/or samples made available by The Children's Brain Tumor Network (formerly the Children's Brain Tumor Tissue Consortium)".

*Accessing the NCI-CDS dataset*

Researchers must create an NIH eRA Commons account (https://public.era.nih.gov) and be classified as a Principal Investigator (or be listed as a collaborating investigator on a PI's application). Using their eRA account, users can log into the dbGaP Authorized Access page (https://dbgap.ncbi.nlm.nih.gov/aa/wga.cgi?page=login) and submit an online Data Access Request (DAR) for the specific dataset with a signed Data Use Certification Agreement. The DAR includes providing basic institutional and contact information including for collaborating users, as well as a research use statement including the objectives of the proposed research, study design, and data use and analysis plan. The NCI Data Access Committee (DAC) will review to confirm that the proposed research is consistent with data use limitations, with data requests to be rejected if there is insufficient detail to make this determination (see tips for DAR preparation here: https://www.ncbi.nlm.nih.gov/projects/gap/cgi-bin/GetPdf.cgi?document_name=GeneralAAInstructions.pdf ). Once approved by the NCIDAC, investigators will then be able to download the data from dbGaP with the Aspera Connect software. Researchers are requested to give annual project updates and include the following acknowledgement statement to reference the usage: "The data from this study phs002517 was made available pre-publication without embargo to support rapid and collaborative research in pediatric cancer via the NCI's Cancer Research Data Commons (https://datacommons.cancer.gov). This availability is made possible with the support of NCI's Childhood Cancer Data Initiative (grant No. 3P30CA082103-21S9) and Gabriella Miller Kids First Pediatric Research Program (X01 CA267587). Initial data generation efforts and coordination costs were supported by a number of philanthropic and industry partners with further details at cbtn.org."

## Code availability

All tools and software packages used in this project are publicly available, including: dcm2niix (version v1.0.20220720)[24], MiDeface (FreeSurfer 7.3.2)[26], CaPTk (1.8.1)[28,29], MRIQC (0.16.1)[33]. The code developed in-house for image de-identification and preparation has been made open source and can be freely accessed at: https://github.com/d3b-center/image-deid-etl.

## Acknowledgements

We would like to express our heartfelt gratitude to the patients and families who have generously donated tumor specimens and data to the CBTN to provide researchers with invaluable resources that enable them to develop better treatment options for children facing this challenging condition. Philanthropic support has ensured the CBTN's ability to collect, store, manage, and distribute specimens and data. The following donors have provided leadership level support for CBTN: CBTN Executive Council members, Brain Tumor Board of Visitors, Children's Brain Tumor Foundation, Easie Family Foundation, Kortney Rose Foundation, Lilabean Foundation, Minnick Family Charitable Fund, Perricelli Family, Psalm 103 Foundation, and Swifty Foundation. We would like to gratefully acknowledge funding support for this project from the NIH trusted partner Leidos (75N91019D00024-75N91020F00003 and 75N91019D00024-75N91021F00013 to A.C.R.), and NIH Office of Data Science Strategy (ODSS) INCLUDE Data Hub supplement (3U2CHL156291-02S1 to A.C.R.).

**Author contributions**

Conceptualization (AMF, AFK, JL, PBS, ACR, AN)

Methodology (AMF, AFK, AN)

Software (AMF, RB, AL)

Analysis (AMF, AFK, HA, KV, NK, SB, DH, SA, RM, AV, JBW, AN)

Data Curation (AMF, AFK, HA, KV, TN, EF, ZH, JH, JS, SC, DM, KD, AW, JM)

Writing - Original Draft (AMF)

Writing - Review & Editing (AMF, AFK, HA, AL, KV, RB, NK, SB, DH, MCK, SA, RM, TQN, EAF, ZH, JH, KF, MGL, UB, YV, JS, SL, RML, SC, AAS, DM, BR, KD, IFP, MW, AV, JBW, SM, PBS, APH, AJW, JVL, JLM, ACR, AN)

Visualization (AMF)

Supervision (ACR, AN)

Resources / funding acquisition (SL, AAS, BR, IFP, RML, APH, SM, AW, JVL, PBS, ACR, AN)

**Competing interests**

The authors have no conflicts of interest to declare.

## Figures

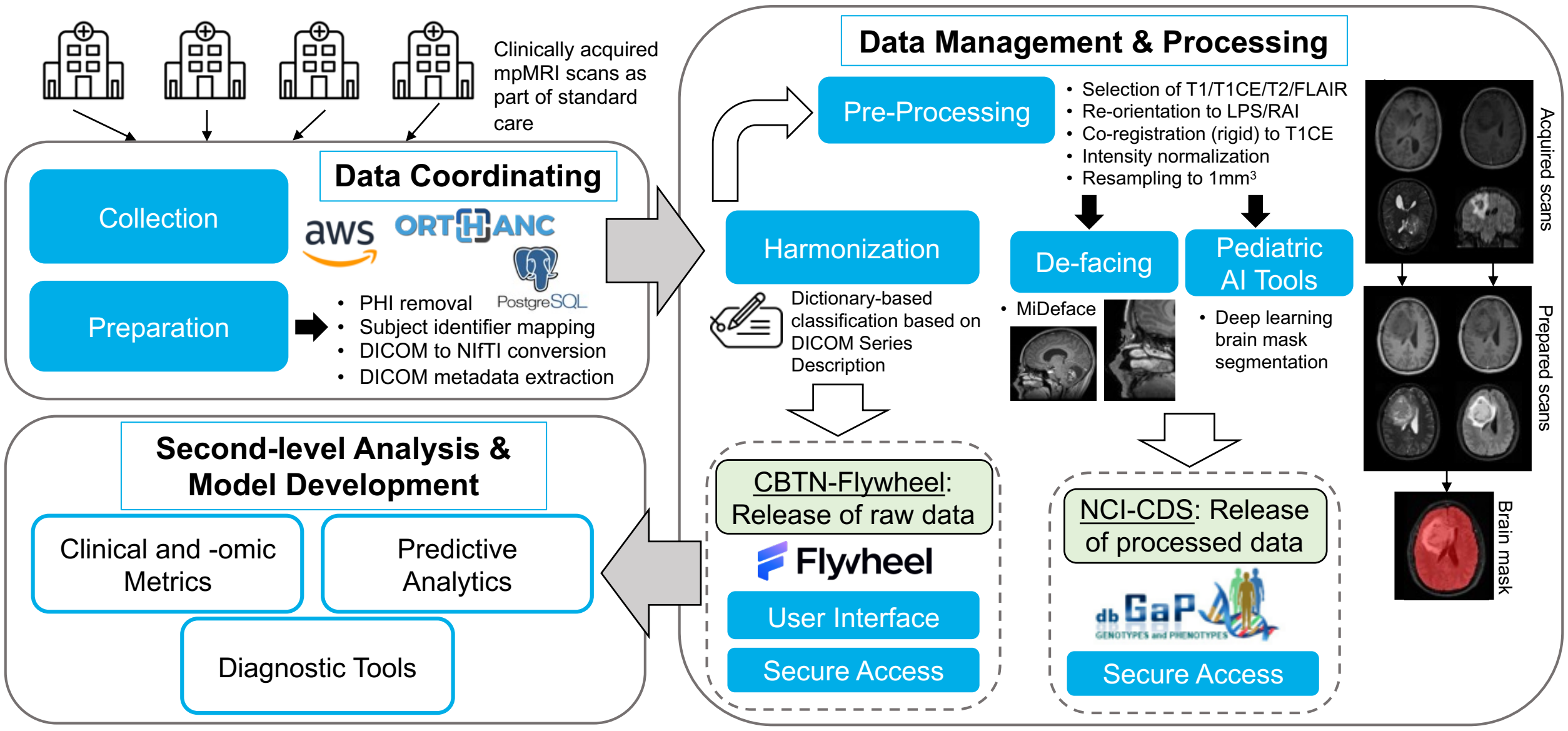

Clinically acquired mpMRI scans as part of standard care
Data Coordinating
Collection
Preparation
aws
ORTHANC
PostgreSQL
• PHI removal
• Subject identifier mapping
• DICOM to NIfTI conversion
• DICOM metadata extraction
Data Management & Processing
Pre-Processing
• Selection of T1/T1CE/T2/FLAIR
• Re-orientation to LPS/RAI
• Co-registration (rigid) to T1CE
• Intensity normalization
• Resampling to 1mm³
Harmonization
Dictionary-based classification based on DICOM Series Description
De-facing
• MiDeface
Pediatric AI Tools
• Deep learning brain mask segmentation
Acquired scans
Prepared scans
Brain mask
CBTN-Flywheel: Release of raw data
Flywheel
User Interface
Secure Access
NCI-CDS: Release of processed data
dbGaP
GENOTYPES and PHENOTYPES
Secure Access
Second-level Analysis & Model Development
Clinical and -omic Metrics
Predictive Analytics
Diagnostic Tools

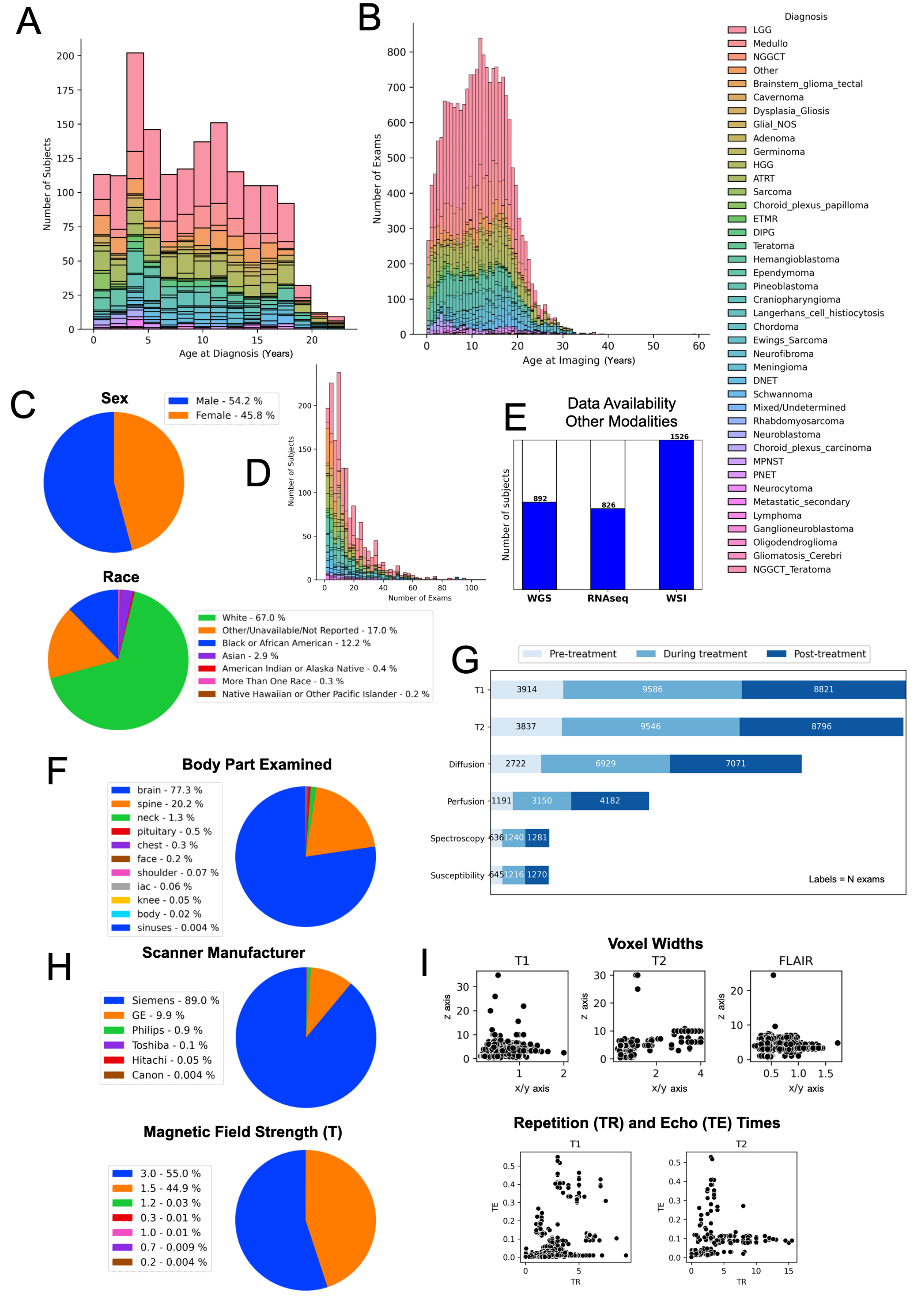
A
Number of Subjects
Age at Diagnosis (Years)
B
Number of Exams
Age at Imaging (Years)
Diagnosis
LGG
Medullo
NGGCT
Other
Brainstem_glioma_tectal
Cavernoma
Dysplasia_Gliosis
Glial_NOS
Adenoma
Germinoma
HGG
ATRT
Sarcoma
Choroid_plexus_papilloma
ETMR
DIPG
Teratoma
Hemangioblastoma
Ependymoma
Pineoblastoma
Craniopharyngioma
Langerhans_cell_histiocytosis
Chordoma
Ewings_Sarcoma
Neurofibroma
Meningioma
DNET
Schwannoma
Mixed/Undetermined
Rhabdomyosarcoma
Neuroblastoma
Choroid_plexus_carcinoma
MPNST
PNET
Neurocytoma
Metastatic_secondary
Lymphoma
Ganglioneuroblastoma
Oligodendroglioma
Gliomatosis_Cerebri
NGGCT_Teratoma
C
Sex
Male - 54.2 %
Female - 45.8 %
Race
White - 67.0 %
Other/Unavailable/Not Reported - 17.0 %
Black or African American - 12.2 %
Asian - 2.9 %
American Indian or Alaska Native - 0.4 %
More Than One Race - 0.3 %
Native Hawaiian or Other Pacific Islander - 0.2 %
D
Number of Subjects
Number of Exams
E
Data Availability
Other Modalities
Number of subjects
892
826
1526
WGS
RNAseq
WSI
F
Body Part Examined
brain - 77.3 %
spine - 20.2 %
neck - 1.3 %
pituitary - 0.5 %
chest - 0.3 %
face - 0.2 %
shoulder - 0.07 %
iac - 0.06 %
knee - 0.05 %
body - 0.02 %
sinuses - 0.004 %
G
Pre-treatment
During treatment
Post-treatment
T1 3914 9586 8821
T2 3837 9546 8796
Diffusion 2722 6929 7071
Perfusion 1191 3150 4182
Spectroscopy 636 1240 1281
Susceptibility 645 1216 1270
Labels = N exams
H
Scanner Manufacturer
Siemens - 89.0 %
GE - 9.9 %
Philips - 0.9 %
Toshiba - 0.1 %
Hitachi - 0.05 %
Canon - 0.004 %
Magnetic Field Strength (T)
3.0 - 55.0 %
1.5 - 44.9 %
1.2 - 0.03 %
0.3 - 0.01 %
1.0 - 0.01 %
0.7 - 0.009 %
0.2 - 0.004 %
I
Voxel Widths
T1
T2
FLAIR
z axis
x/y axis
Repetition (TR) and Echo (TE) Times
T1
T2
TE
TR

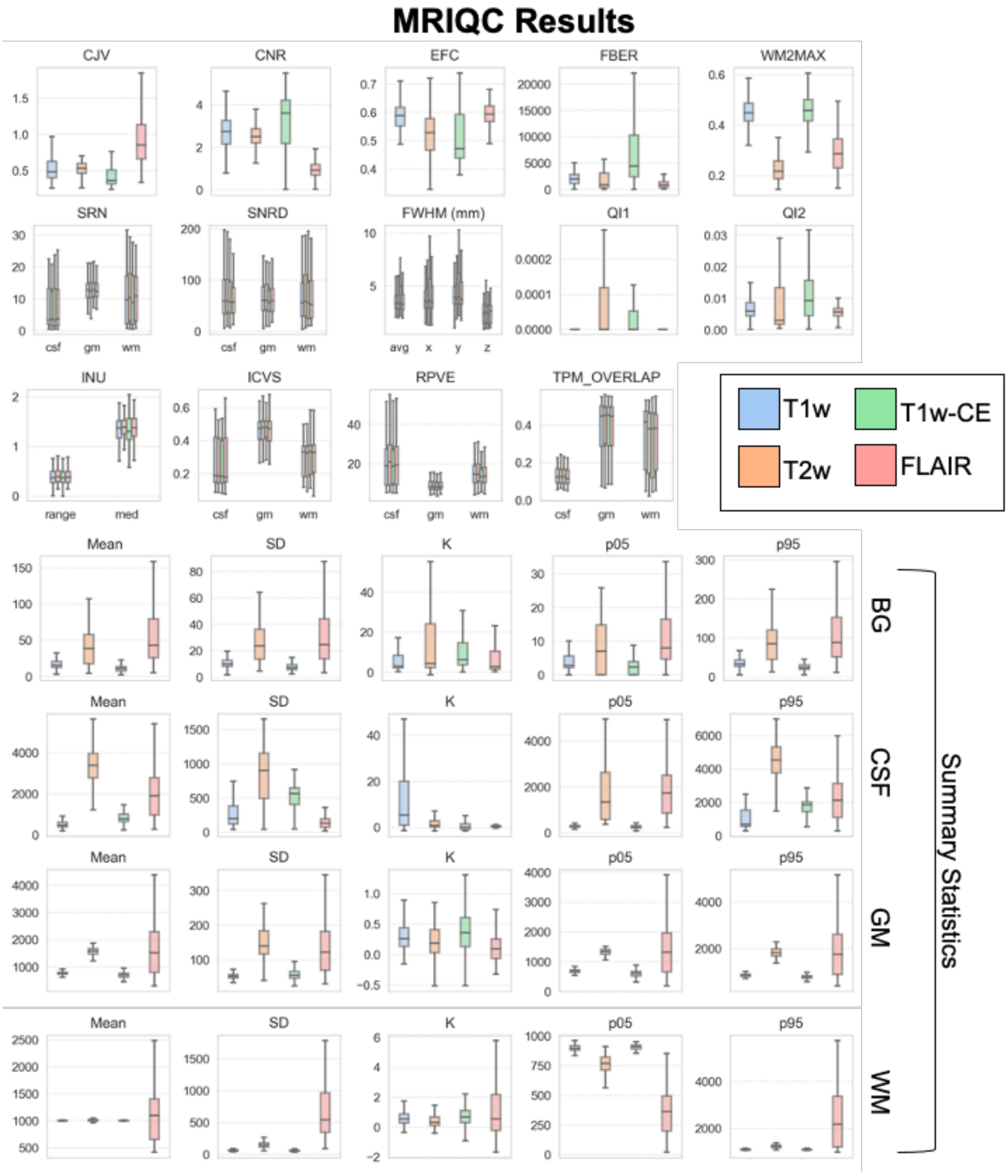

## Figure Legends

**Figure 1. Diagram of overall workflow.** Clinical imaging exams are received from various institutions and centralized. Exams are de-identified of PHI and metadata is warehoused using cloud services and in-house software packages. Images (NIfTIs with JSON sidecar) are uploaded to the Flywheel platform where they are stored and shared with the broader research community (CBTN-Flywheel data). Standard tools are utilized to prepare selected images for research use. Existing standardized pipelines and pediatric-specific AI-powered brain extraction processes are used to generate analysis-ready data files. Processed images and brain mask segmentations are stored and shared as part of a CBTN study in the NCI's Cancer Data Service via the database of Genotypes and Phenotypes (dbGaP; NCI-CDS data).

**Figure 2. CBTN radiology dataset characteristics.** Distributions of CBTN imaging data by cancer diagnosis: (A) histogram of the number of subjects with imaging data by subject age (years) at first diagnosis; (B) histogram of the number of imaging exams by the age (years) at the time of imaging; (D) number of exams per subject. (C) Patient-level demographics of sex and race. (E) Number of subjects with associated tumor specimen collection with available whole genome sequencing (WGS), RNA sequencing (RNAseq), and digital histopathology slides (whole slide imaging, WSI). Exam-level distributions, including: (F) percentage of exams by each body part examined; (G) number of exams for most frequent image types (T1w, T2w, diffusion, perfusion, spectroscopy, susceptibility) categorized by acquisition time relative to treatment status (acquired before, during, or after treatment); (H) percentage of exams by scanner manufacturer and magnetic field strength. (I) Image-level distributions of voxel width along

sagittal/coronal (x/y; x-axis) and transverse (z; y-axis) planes for T1w, T2w, and FLAIR images; and repetition time (TR; x-axis) and echo time (TE; y-axis) for all T1w and T2w images. Each point represents one image.

**Figure 3. MRIQC results for selected cohort.** Distributions of Image Quality Metrics (IQMs) for T1w, T2w, T1w-CE, and FLAIR images based on the unprocessed images of the subset cohort (N=150). Summary statistics shown in bottom rows for background (BG), CSF, white matter (WM), and gray matter (GM) regions. Error bars represent 95% confidence interval. See Table 2 for IQM definitions.

## Tables

| Release information | CBTN-Flywheel (unprocessed data) | NCI-CDS (processed data) |
|---|---|---|
| Managing organization & hosting platform | CBTN Flywheel | NCI CDS via dbGaP |
| Access – requirements | CBTN research project approval; Signed CBTN DUA | NCI dbGaP Data Access Request approval; eRA with Primary Investigator status; Signed NCI DUA |
| Access – getting started resources | https://redcap.chop.edu/surveys/?s=A7M873HMN8 | https://www.ncbi.nlm.nih.gov/projects/gap/cgi-bin/study.cgi?study_id=phs002517.v1.p1 |
| File format | Compressed NIfTI (.nii.gz) with paired JSON (additional DICOM metadata) | Compressed NIfTI (.nii.gz) |
| File names | DICOM Series Description tag | [subID]_[age]_[T1/T1CE/T2/FL/pred_brainMask]_to_SRI_defaced.nii.gz |
| De-faced (face regions removed) | No | Yes |
| **Patient and Scan Characteristics** | | |
| Total Patients | 1,526 | 370 |
| Total Sessions (number of exams)<br>Treatment-naïve<br>During treatment<br>After treatment<br>Treatment information unknown | 23,101<br>4,083<br>9,941<br>9,051<br>26 | 370<br>370 |
| Age at diagnosis, range (years)<br>Age at diagnosis, median (years) | 0.003 – 20.96<br>8.81 | 0.24 – 20.73<br>8.43 |
| Legal Sex<br>Male<br>Female<br>Unknown | <br>826<br>698<br>2 | <br>182<br>188 |
| Overall Survival<br>Range (years)<br>Median (years)<br>Unknown (N) | <br>0.003 to 118.45<br>3.53<br>131 | <br>0.06 to 21.21<br>3.66<br>8 |
| Event Free Survival<br>Range (years)<br>Median (years)<br>Unknown (N) | <br>0.003 to 116.8<br>1.94<br>133 | <br>0.04 to 13.29<br>2.08<br>8 |
| Race (N) | | |
| White | 1,021 | 269 |
| Black or African American | 186 | 40 |
| Asian | 44 | 7 |
| Native Hawaiian or Other Pacific | 3 | 2 |
| Islander | 6 | 3 |
| American Indian or Alaska Native | 5 | 1 |
| More than one race | 261 | 48 |
| Information unavailable | | |
| Associated subject-level data availability | | |
| (N) | 892 | 261 |
| Whole Genome Sequencing | 826 | 255 |
| RNAseq | 1,526 | 370 |
| Digital pathology slides | | |
| Histology (N subjects) | | |
| Low Grade Glioma / astrocytoma | 519 | 208 |
| Medulloblastoma | 159 | 99 |
| High Grade Glioma / astrocytoma | 138 | 29 |
| Other | 116 | |
| Ependymoma | 89 | 8 |
| Craniopharyngioma | 71 | 1 |
| Atypical Teratoid Rhabdoid (ATRT) | 46 | |
| Meningioma | 38 | |
| Germinoma | 36 | 2 |
| Neurofibroma | 36 | |
| DIPG | 29 | 18 |
| DNET | 29 | |
| Schwannoma | 28 | |
| Glial_NOS | 27 | |
| Choroid_plexus_papilloma | 23 | |
| Cavernoma | 20 | |
| Ewings_Sarcoma | 19 | |
| Neuroblastoma | 14 | |
| Teratoma | 13 | 1 |
| PNET | 13 | 1 |
| Sarcoma | 11 | |
| Pineoblastoma | 11 | |
| Adenoma | 10 | |
| Langerhans_cell_histiocytosis | 10 | |
| Metastatic_secondary | 10 | |
| NGGCT | 9 | |
| Hemangioblastoma | 9 | |
| ETMR | 7 | 2 |
| MPNST | 7 | |
| Chordoma | 6 | |
| Neurocytoma | 6 | 1 |
| Choroid_plexus_carcinoma | 4 | |
| Oligodendroglioma | 3 | |
| Brainstem_glioma_tectal | 2 | |
| Rhabdomyosarcoma | 2 | |
| Lymphoma | 2 | |
| Gliomatosis_Cerebri | 2 | |
| Ganglioneuroblastoma | 1 | |
| Multiple | 47 | |
| Scanner Magnetic Field Strength (T) | | |
| 3 | 12,415 | 224 |
| 1.5 | 10,127 | 144 |
| 1.2 | 7 | |
| 1.0 | 3 | |
| 0.7 | 2 | 1 |
| 0.3 | 3 | |
| Unknown | 543 | 1 |
| Scanner Manufacturer | | |
| Siemens | 20,256 | 308 |
| GE | 2,257 | 49 |
| Phillips | 208 | 11 |
| Toshiba | 34 | 2 |
| Hitachi | 12 | |
| Canon | 1 | |
| Unknown | 333 | |

**Table 1.** Release information, patient, and exam characteristics for the CBTN-Flywheel full dataset (center column) and NCI-CDS selected pre-processed cohort (right column).

| | ***Benchmark*** | ***T1w Median (SEM)*** | ***T2w Median (SEM)*** | ***Measure*** |
|---|---|---|---|---|
| ***Coefficient of joint variation (CJV)*** | T1w: 0.26 – 1.01*<br>T2w: 0.2 – 4.27*<br><br>Lower values are better. | 0.48 (.04) | 0.53 (.05) | Gray-to-white matter contrasts; can detect heavy head motion and large INU artifacts |
| ***Entropy focus criterion (EFC)*** | T1w: 0.44 – 0.73*<br>T2w: 0.45 – 0.95*<br><br>Lower values are better. | 0.59 (.004) | 0.53 (.007) | Entropy of voxel intensities (EFC=0 is all energy concentrated in one voxel[41]); can detect ghosting and blurring induced by head motion |
| ***Intensity non-uniformity (INU)*** | Small spreads around 1.0 | 1.5 (.02) | 1.21 (.01) | Location and spread of bias field estimated by INU correction[42] |
| ***Mortamet's quality index 1 (QI1)*** | <.01 (lower values are better) | 0 (.003) | 0 (.003) | Percent of voxels in background with intensities corrupted by artifact[43] |
| ***Mortamet's quality index 2 (QI2)*** | <.05 (lower values are better) | 0.006 (.001) | 0.003 (.003) | Noise intensity distribution after correction of QI1[43] |
| ***Foreground-to-background energy ratio (FBER)*** | T1w: 152 – 2042*<br>T2w: 536 – 3668*<br><br>Higher values better. | 1968 (754) | 882 (697) | Mean energy of image values within head relative to background |
| ***Voxel smoothness (FWHM)*** | T1w: 1.27 – 5.51*<br>T2w: 2.11 – 4.31*<br><br>Lower values better. | 3.62 (.06) | 3.79 (.11) | Global image blur (in mm) |
| ***Signal-to-noise ratio (SNR)*** | T1w: 5.2 – 16.3*<br>T2w: 3.4 – 14.6*<br><br>Higher values better. | 11.3 (.17) | 7.6 (.15) | Signal intensity in tissue compared to background |
| ***Contrast-to-noise ratio (CNR)*** | T1w: 1.18 – 4.67*<br>T2w: 0.3 – 4.13*<br><br>Higher values better. | 2.7 (.07) | 2.5 (.08) | Separation of gray and white matter distributions |
| ***White matter to maximum intensity ratio (WM2MAX)*** | Around 0.6-0.8<br><br>T1w: 0.28 – 0.8*<br>T2w: 0.14 – 0.88* | 0.45 (.01) | 0.22 (.01) | Median intensity within WM mask compared to full intensity distribution; can detect long tails due to hyper-intensity |
| ***Intracranial volume estimation (ICVs)*** | CSF: 0.2<br>WM: 0.45<br>GM: 0.35<br><br>T1w:<br>CSF: 0.13 – 0.28*<br>GM: 0.35 – 0.52*<br>WM: 0.33 – 0.43*<br>T2w:<br>CSF: 0.11 – 0.51*<br>GM: 0.04 – 0.66*<br>WM: 0.05 – 0.69* | CSF: 0.17 (.003)<br>WM: 0.35 (.004)<br>GM: 0.48 (.01) | CSF: 0.15 (.008)<br>WM: 0.35 (.007)<br>GM: 0.50 (.01) | Percent volume of each tissue type |
| ***Residual partial volume effect (rPVE)*** | T1w:<br>CSF: 7.9 – 62.8*<br>GM: 5.4 – 26.7*<br>WM: 7.1 – 35.1*<br>T2w:<br>CSF: 12.4 – 52.2*<br>GM: 5.9 – 164.6*<br>WM: 5.4 – 112.1*<br><br>Lower values better. | CSF: 21.1 (.5)<br>WM: 11.1 (.3)<br>GM: 8 (.2) | CSF: 22.5 (1)<br>WM: 10.9 (.8)<br>GM: 7.9 (.5) | Tissue-wise sum of partial volumes (proportion of tissues present in a voxel), indicating separation of the tissue classes |
| ***Tissue probability maps (TPMs)*** | T1w:<br>CSF: 0.09 – 0.25*<br>GM: 0.3 – 0.55*<br>WM: 0.28 – 0.59*<br>T2w:<br>CSF: 0.05 – 0.21*<br>GM: 0.03 – 0.53*<br>WM: 0.04 – 0.55*<br><br>Higher values indicate greater overlap. | CSF: 0.17 (.003)<br>WM: 0.46 (.01)<br>GM: 0.48 (.01) | CSF: 0.14 (.003)<br>WM: 0.41 (.01)<br>GM: 0.47 (.01) | Overlap between maps estimated from the image and maps based on population atlas (ICBM 2009c template) |

**Table 2. MRIQC results for T1w and T2w images in the pre-processed cohort (N = 150).**

Image Quality Metrics (IQMs) provided generated by MRIQC for the unprocessed images of 150 pediatric brain tumor subjects. Benchmark value ranges marked with an asterix (*) are the $5^{th}$ - $95^{th}$ percentiles from the MRIQC crowdsourced database with 51,113 T1w samples and 767 T2w samples (calculated based on data from “T1w_curated.csv” and “T2w_curated.csv” provided in prior study).[40] Otherwise the benchmarks reflect those provided in MRIQC documentation and/or original methodology papers.